\documentclass[12pt]{article} 

\usepackage{amsmath}
\usepackage{graphicx}
\usepackage{epsfig}

\usepackage[utf8]{inputenc}    
\usepackage[T1]{fontenc}

\newcommand{\noi}{\noindent}
\newcommand{\beq}{\begin{equation}}
\newcommand{\eeq}{\end{equation}}
\newcommand{\bea}{\begin{eqnarray}}
\newcommand{\eea}{\end{eqnarray}}
\newcommand{\ds}{\displaystyle}

\title{ \bf \huge Gluon propagators in QC$_2$D at high baryon density}

\author{\Large Vitaly Bornyakov $^{1,2,3}$, Andrey Kotov $^{2,4}$, Alexander Nikolaev$^{5}$\\ \Large and Roman Rogalyov $^{1}$}

\begin{document}

\maketitle
\thispagestyle{empty}
\begin{abstract}
We study the transverse and longitudinal gluon propagators in the Landau-gauge lattice
QCD with gauge group $SU(2)$  at nonzero quark chemical potential and zero temperature.
We show that both propagators demonstrate substantial dependence on the quark chemical potential. 
This observation contradicts to earlier findings by other groups.
\end{abstract}
 
\vspace*{35mm}

\noindent $^{1}$ \quad NRC “Kurchatov Institute” - IHEP, 142281 Protvino, Russia\\
$^{2}$ \quad NRC “Kurchatov Institute” - ITEP, Moscow, 117218 Russia\\
$^{3}$ \quad   Far Eastern Federal University, School of Biomedicine, 690950 Vladivostok, Russia\\
$^{4}$ \quad  Bogoliubov Laboratory of Theoretical Physics, Joint Institute for Nuclear Research, Dubna, 141980 Russia\\
$^{5}$ \quad  Department of Physics, College of Science, Swansea University, Swansea SA2 8PP,
United Kingdom




\newpage 

\section{Introduction}
\label{sec:introduction}

The properties of nuclear matter at low temperature and high density and the location of the phase transition to deconfined quark matter are subjects of both experimental and theoretical studies.
It is known that the non-perturbative first principles approach as lattice QCD is  inapplicable at large baryon densities and small temperatures due to the so-called sign problem.
This makes important to study the QCD-like models \cite{Kogut:2000ek}, in particular lattice $SU(2)$ QCD (also called QC$_2$D). The properties of this theory were studied with the help of various approaches: chiral perturbation theory \cite{Kogut:2000ek,Splittorff:2001fy,Kanazawa:2009ks}, Nambu-Jona-Lasinio model \cite{Brauner:2009gu,Sun:2007fc,He:2010nb}, quark-meson-diquark model \cite{Strodthoff:2011tz,Strodthoff:2013cua}, random matrix theory \cite{Vanderheyden:2001gx,Kanazawa:2011tt}. 
Supported by agreement with high precision lattice results obtained in SU(2) QCD these methods can be applied to real QCD with higher confidence. Lattice studies were made with staggered fermions \cite{Hands:1999md,Kogut:2001if,Kogut:2001na,Kogut:2002cm,Braguta:2016cpw,Bornyakov:2017txe,Astrakhantsev:2018uzd,Wilhelm:2019fvp} for $N_f=4$ or, more recently, $N_f=2$  and Wilson fermions \cite{Nakamura:1984uz,Hands:2006ve,Hands:2010gd,Hands:2011ye,Cotter:2012mb,Boz:2018crd} for $N_f=2$ mostly. 

The phase structure of $N_f=2$ QC$_2$D at large baryon density and $T=0$ was studied recently in \cite{Bornyakov:2017txe}. The simulations were carried out at small lattice spacing and the range of large quark chemical potential was reached without strong lattice artifacts. The main result of Ref.~\cite{Bornyakov:2017txe} is the observation that the string tension $\sigma$ is compatible with zero for $\mu_q$ above 850~MeV. It was also found that the so called spatial string tension $\sigma_s$ started to decrease at approximately same value of $\mu_q$ and went to zero at $\mu_q > 2000$~MeV.

The gluon propagators are among important quantities, e.g. they play crucial role in the Dyson-Schwinger equations approach.  
In this paper we present results of our study of dependence of the gluon propagators and respective screening masses on $\mu_q$, including large $\mu_q$ values range. We also look for signals of the confinement-deconfinement transition in the propagators behavior.

Landau gauge gluon propagators were extensively studied in the infrared range of momenta by
various methods. We shall note lattice gauge theory, Dyson-Schwinger equations, Gribov-Zwanziger
approach. At the same time the studies in the particular case of nonzero quark chemical potential are restricted to a few papers only. For the lattice QCD this is explained by the sign problem mentioned above. 

The gluon propagators in lattice QC$_2$D were recently studied for the first time in Ref.~\cite{Boz:2018crd}. 
The main conclusion of Ref.~\cite{Boz:2018crd} was that the gluon propagators practically do not change for the range of $\mu_q$ values studied: $\mu_q < 1.1$~GeV.  Our main conclusion is opposite. We found
substantial influence of the quark chemical potential on the gluon propagators starting from rather low values ($\mu_q \sim 300$ MeV) and increasing with increasing $\mu_q$.
Thus results presented in Ref.~\cite{Boz:2018crd} differ from our results presented in this paper in many respects.  The reason for these rather drastic differences might be that the lattice action and lattice spacing differ from those used in our study. 

\section{Lattice setup}

For numerical simulations we used the tree level improved Symanzik gauge action~\cite{Weisz:1982zw}
and the staggered fermion action of the form
\begin{equation}
\label{eq:S_F}
S_F = \sum_{x, y} \bar \psi_x M(\mu, m)_{x, y} \psi_y + \frac{\lambda}{2} \sum_{x} \left( \psi_x^T \tau_2 \psi_x + \bar \psi_x \tau_2 \bar \psi_x^T \right)
\end{equation}
with
\begin{equation}
\label{eq:Dirac_operator}
M(\mu,m)_{xy} = ma\delta_{xy} + \frac{1}{2}\sum_{\nu = 1}^4 \eta_{\nu}(x)\Bigl[ U_{x, \nu}\delta_{x + h_{\nu}, y}e^{\mu a\delta_{\nu, 4}} - U^\dagger_{x - h_{\nu}, \nu}\delta_{x - h_{\nu}, y}e^{- \mu a\delta_{\nu, 4}} \Bigr]\,,
\end{equation}
where $\bar \psi$, $\psi$ are staggered fermion fields, $a$ is the lattice spacing, $m$ is the bare quark mass, and $\eta_{\nu}(x)$ are the standard staggered phase factors.
The quark chemical potential $\mu$ is introduced into the Dirac 
operator~\ref{eq:Dirac_operator} through the multiplication of the lattice gauge 
field components $U(x,4)$ and $U^\dagger(x,4)$ by factors $e^{\pm \mu a}$, respectively. 

We have to add to the standard staggered fermion action a diquark source term~\cite{Hands:1999md}. This term explicitly violates $U_V(1)$ symmetry and allows to observe diquark 
condensation on finite lattices, because this term effectively chooses one vacuum from the family of $U_V(1)$-symmetric vacua. 
Typically one carries out numerical simulations at a few nonzero values of the parameter $\lambda$ and then extrapolates to $\lambda=0$. 
The  lattice configurations we are using are generated at one small  value $\lambda=0.00075$ which is much smaller than the quark mass in lattice units. 

Integrating out the fermion fields, the partition function for the theory with the action $S=S_G+S_F$
can be written in the form
\beq
Z = \int DU e^{-S_G} \cdot Pf \begin{pmatrix} \lambda \tau_2  & M \\ - M^T & \lambda \tau_2 \end{pmatrix} = \int DU e^{-S_G} \cdot {\bigl ( \det (M^\dagger M + \lambda^2) \bigr )}^{\frac 1 2},
\label{z1}
\eeq
which corresponds to $N_f=4$ dynamical fermions in the continuum limit. Note that the pfaffian $Pf$ is strictly positive, such that one can use 
Hybrid Monte-Carlo methods to compute the integral.
First lattice studies of the theory with partition function (\ref{z1}) have been carried 
out in Refs.~\cite{Kogut:2001na, Kogut:2001if, Kogut:2002cm}. We study a theory with the partition function
\beq
Z=\int DU e^{-S_G} \cdot {\bigl ( \det (M^\dagger M + \lambda^2) \bigr )}^{\frac 1 4},
\label{z2}
\eeq
corresponding to $N_f=2$ dynamical fermions in the continuum limit.

We are using lattice gauge field configurations generated in 
Ref.~\cite{Bornyakov:2017txe} on $32^4$ lattices for a set of the chemical potentials in the range $a\mu_q \in (0, 0.3)$.
At zero density scale was set using the QCD Sommer scale value $r_0=0.468(4) \mathrm{~fm}$ ~\cite{Bazavov:2011nk}.
We found \cite{Bornyakov:2017txe}  $r_0/a=10.6(2)$. Thus lattice spacing is $a = 0.044(1) \mathrm{~fm}$ while the string tension at $\mu_q=0$ is $\sqrt{\sigma_0}=476(5) \mathrm{~MeV}$.  
The pion is rather heavy with its mass $m_{\pi}=740(40) \mathrm{~MeV}$.

We employ the standard definition of the lattice gauge vector
potential $A_{x,\mu}$ \cite{Mandula:1987rh}:
\beq
A_{x,\mu} = \frac{Z}{2iag}~\Bigl( U_{x\mu}-U_{x\mu}^{\dagger}\Bigr)
\equiv A_{x,\mu}^a \frac{\sigma_a}{2} \,,
\label{eq:a_field}
\eeq
where $Z$ is the renormalization factor.
The lattice Landau gauge fixing condition is
\beq
(\nabla^B A)_{x} \equiv {1\over a} \sum_{\mu=1}^4 \left( A_{x,\mu}
- A_{x-a\hat{\mu},\mu} \right)  = 0 \; ,
\label{eq:diff_gaugecondition}
\eeq

\noi which is equivalent to finding an extremum of the gauge functional

\beq
F_U(\omega)\; =\; \frac{1}{4V} \sum_{x\mu}\ \frac{1}{2} \ T\!r\; U^{\omega}_{x\mu} \;,
\label{eq:gaugefunctional}
\eeq

\noi with respect to gauge transformations $\omega_x~$.  
To fix the Landau gauge we use the simulated annealing (SA) algorithm
with finalizing overrelaxation \cite{Bornyakov:2009ug}. 
To estimate the Gribov copy effect,
we employed five gauge copies of each configuration; however, the difference between the "best-copy" and "worst-copy" values of each quantity under consideration lies within statistical errors.

The gluon propagator $D_{\mu\nu}^{ab}(p)$ is defined
as follows:
\beq
D_{\mu\nu}^{ab}(p) = \frac{1}{Va^4}
    \langle \widetilde{A}_{\mu}^a(q) \widetilde{A}_{\nu}^b(-q) \rangle\;,
\qquad \mbox{where} \qquad
\widetilde A_\mu^b(q) = a^4 \sum_{x} A_{x,\mu}^b
\exp\Big(\ iq(x+{\hat \mu a\over 2}) \Big),
\label{eq:gluonpropagator}
\eeq
$q_i \in (-N_s/2,N_s/2]$, $ q_4 \in (-N_t/2,N_t/2]$  and 
the physical momenta $p_\mu$ are defined by the relations $ap_{i}=2 \sin{(\pi q_i/N_s)}$,
$ap_{4}=2\sin{(\pi q_4/N_t)}$.

At nonzero $\mu_q$ the $O(4)$ symmetry is broken and 
there are two tensor structures for the gluon propagator \cite{Kapusta:2006pm}~:
\beq
D_{\mu\nu}^{ab}(p)=\delta_{ab} \left( P^T_{\mu\nu}(p)D_{T}(p) +
P^L_{\mu\nu}(p)D_{L}(p)\right)\,.
\eeq

In what follows we consider the  softest mode $p_4=0$ 
and use the notation $p=|\vec p|$ and $D_{L,T}(p)=D_{L,T}(0,|\vec{p}|)$. 
In this case, (symmetric) orthogonal projectors $P^{T;L}_{\mu\nu}(p)$
are defined as follows:
\beq
P^T_{ij}(p)=\left(\delta_{ij} - \frac{p_i p_j}{\vec{p}^2} \right),\,
~~~P^T_{\mu 4}(p)=0~;~~~~~P^L_{44}(p) = 1~;~~P^L_{\mu i}(p) = 0 \,.
\eeq
\noi Therefore, two scalar propagators - longitudinal $D_{L}(p)$ and
transverse $D_T(p)$ -  are given by

\begin{displaymath}  \label{gluonpropagator}
D_T(p)= \left\{ \begin{array}{ll}
\frac{1}{6} \sum_{a=1}^{3}\sum_{i=1}^{3} D_{ii}^{aa}(p) & \mbox{if} \quad p\neq 0~  \\
\frac{1}{9} \sum_{a=1}^{3}\sum_{i=1}^{3} D^{aa}_{ii}(p) & \mbox{if} \quad p=0~ 
\end{array}\right.\ , \qquad 
D_L(p)= \frac{1}{3}\sum_{a=1}^{3} D_{44}^{aa}(p) \,, \nonumber \\
\end{displaymath}

$D_T(p)$  is associated with the magnetic sector, $D_L(p)$ -- with the electric sector.

\section{Gluon propagators and screening masses}

We begin with the analysis of the propagators in the infrared domain
where their behavior is conventionally described in terms of the so called 
screening masses.

\subsection{Definition of the screening mass}
In the studies of the gluon propagators at finite temperatures/densities
two definitions of the gluon screening mass are widely used. 
The first definition is as follows: chromoelectric(magnetic) screening mass 
is the parameter $\tilde{m}$ that appears in the Taylor expansion of 
the respective (longitudinal or transverse) propagator 
at zero momentum (see Refs. \cite{Bornyakov:2010nc,Dudal:2018cli})
\beq\label{scrmass_mom1}
D_{L,T}^{-1}(p) = \zeta (\tilde{m}_{E,M}^2 + p^2 + \overline{o}(p^2))\; .
\eeq
The second one was proposed by A.Linde \cite{Linde:1980ts}
for high orders of finite-temperature perturbation theory
to make sense, it has the form
\beq
 m_M^2={1\over D_T(p=0)}\;.
\eeq
Analogous quantity in the chromoelectric sector 
\beq
m_E^2={1\over D_L(p=0)} 
\eeq
is often referred to as the chromoelectric screening mass \cite{Maas:2011se}.
These masses can be related by the factor $\zeta$,
\beq
m_{E,M}^2 = \zeta \tilde{m}_{E,M}^2\;.
\eeq
If $\zeta$ is independent of the thermodynamical parameters, two definitions
can be considered as equivalent (they differ by a constant factor and 
thermodynamical information is contained only in the dependence on the parameters). 
However, this is not always the case. To discriminate between them,
we will label the former mass $\tilde m_{E,M}$ as the 
proper screening mass and the latter $m_{E,M}$ as the Linde screening mass.

We consider both masses in our study. Similar approach was considered in \cite{Dudal:2018cli}.

\subsection{Screening masses in QC$_2$D}

\begin{figure}[tbh]
\vspace*{-17mm}\hspace*{-0.6cm}
\includegraphics[width=8.2cm]{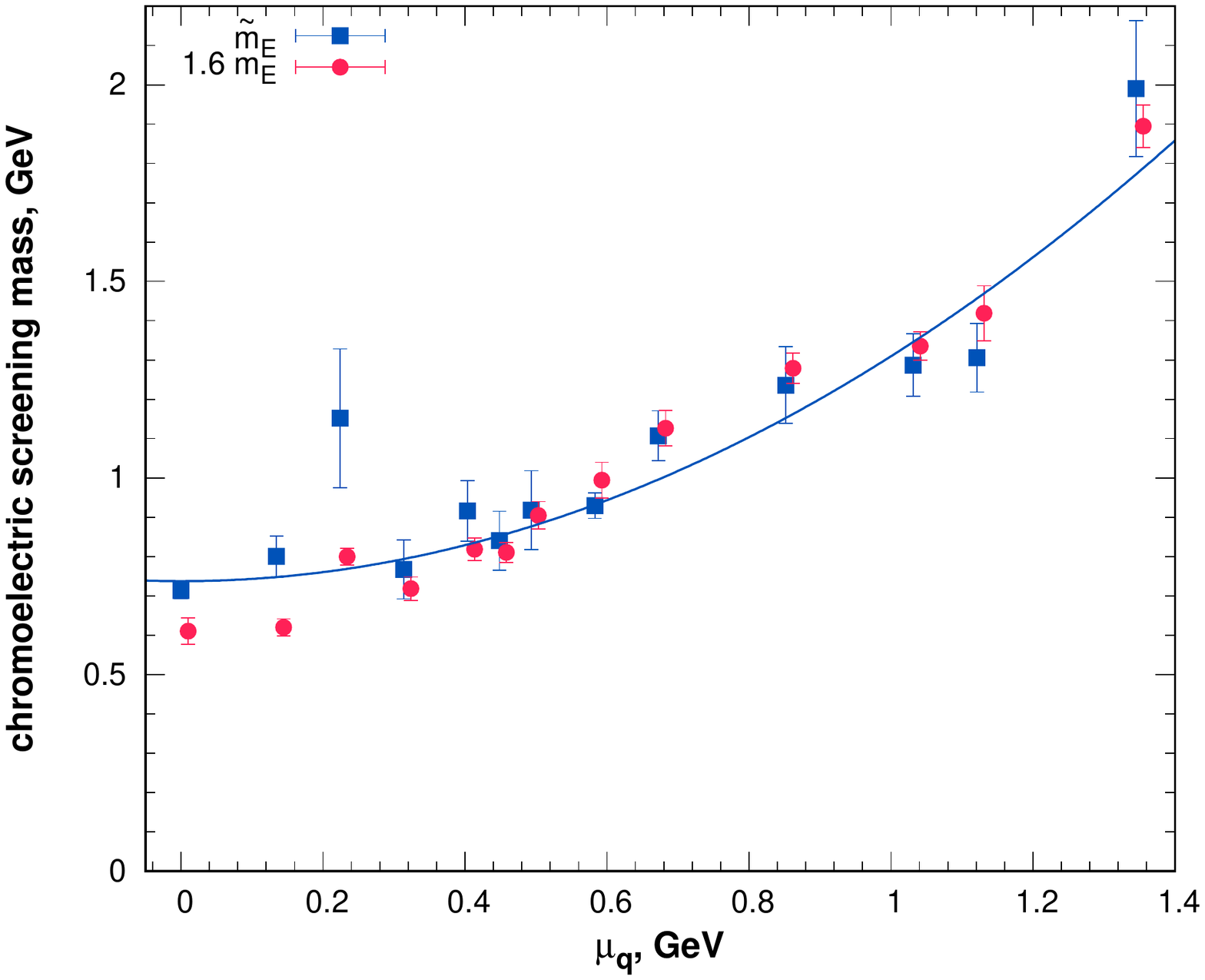}
\includegraphics[width=8.2cm]{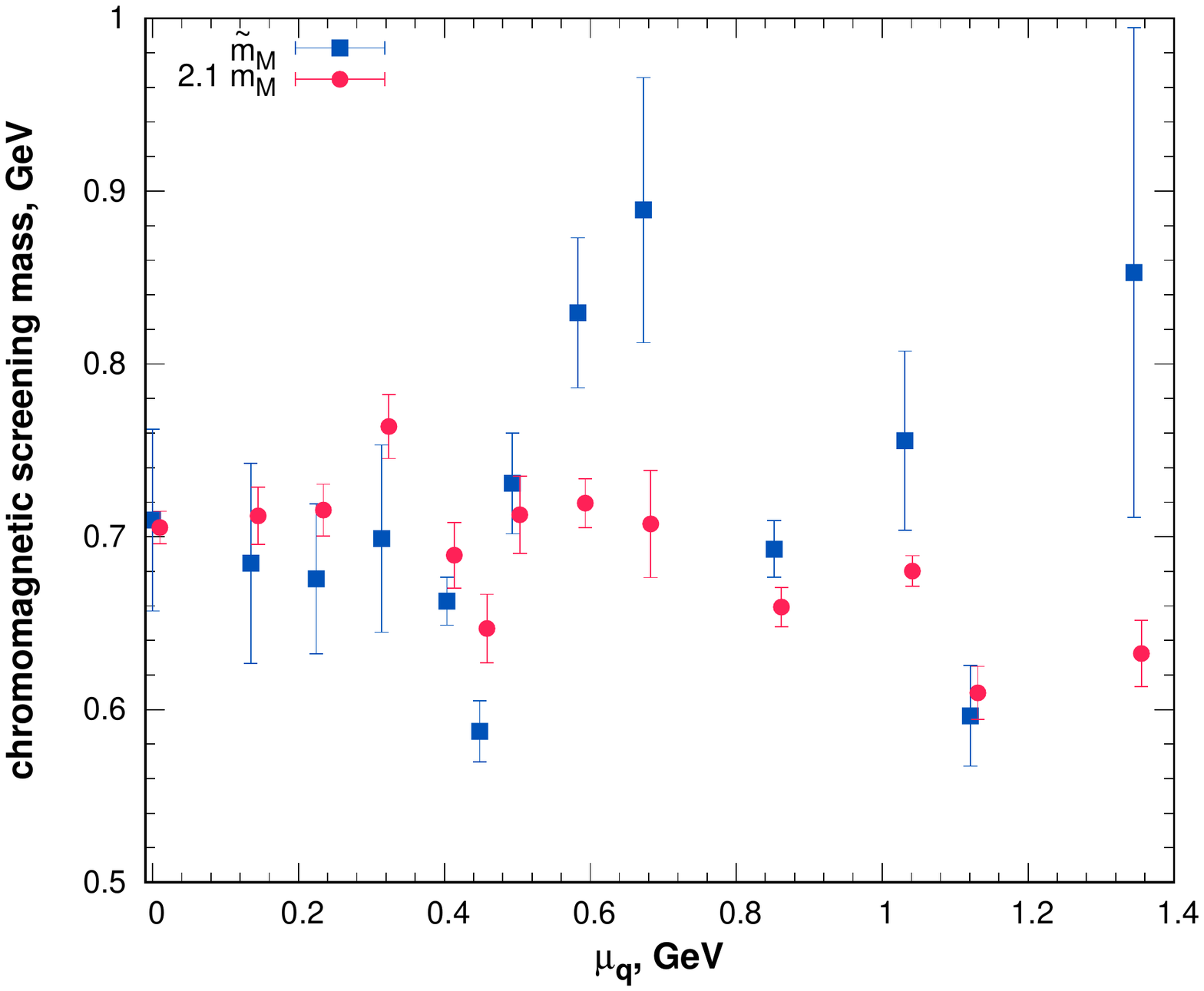}
\vspace*{-29mm}
\caption{
Chromoelectric (left) and chromomagnetic (right) 
Linde and proper screening masses as functions of $\mu_q$.
The Linde mass $m_E$ is obtained from the propagators normalized at 6~GeV;
to compare its dependence on $\mu_q$ with that of $\tilde{m}_E$,
we show $1.6 m_E$  for the chromoelectric mass 
and $2.1 m_M$ for the chromomagnetic mass.}
\label{fig:both_proper_n_Linde_GeV}
\end{figure}

We make fits over the extended range of momenta $p < p_{cut}=2.3$ GeV, 
comparatively high momenta are allowed for 
because our minimal momentum is as big as  $p_{min}=0.88$~GeV.

We use the fit function 
\beq \label{eq:LOW_MOM_fit_fun_E}
D_{L}^{-1}(p) = \zeta_E (\tilde{m}_{E}^2 + p^2 + r_E\;p^4) 
\eeq
for the chromoelectric sector and
\beq \label{eq:LOW_MOM_fit_fun_M}
D_{T}^{-1}(p) = \zeta_M (\tilde{m}_{M}^2 + p^2 + r_M\; p^4 + s_M\; p^6 ) 
\eeq
for the cromomagneic sector.
These fit functions and the cutoff momentum $p_{cut}=2.3$~GeV are chosen 
for the following reasons:
(i) fit function of the type (\ref{eq:LOW_MOM_fit_fun_E}) 
does not work for the transverse propagator: goodness-of-fit is not acceptable (typical $p$-value is of order $10^{-5}$); (ii) it is unreasonable to use fit function of the type (\ref{eq:LOW_MOM_fit_fun_M}) in the chromoelectric sector because the parameters in this function 
are poorly determined, whereas satisfactory goodness-of-fit can be achieved with the 3-parameter fit; 
(iii) higher values of $p_{cut}$ results in a decrease of goodness-of-fit, whereas lower values result in large errors in the parameters, 
however, at $\mu<0.3$~GeV in the chromoelectric sector 
this is not the case and we choose\footnote{An 
important argument for this choice is that the perturbative domain
in the chromoelectric sector at $\mu<0.3$~GeV involves momenta $\simeq 2$~GeV, see below.}
$p_{cut}=1.8$~GeV.

We checked stability of the proper chromoelectric screening mass 
against an exclusion of zero momentum from our fit domain. 
At $\mu_q < 0.3$~GeV this procedure results in an increase 
of $\tilde{m}_E$ by more than two standard deviations, 
whereas at higher $\mu_q$ the value of $\tilde{m}_E$
changes within statistical errors. 

As for the chromomagnetic screening mass, an exclusion of zero momentum
results in a dramatic increase of its uncertainty. Thus the longitudinal propagator
considered over the momentum range $0.8<p<2.3$~GeV does involve an information 
on the respective screening mass, whereas the transverse propagator --- does not.

The dependence of both $\tilde{m}_{E}$ and $m_{E}$
on the quark chemical potential is depicted in Fig.\ref{fig:both_proper_n_Linde_GeV}, left panel.
It is seen that at $\mu_q<0.3$~GeV the difference between $\tilde{m}_{E}$ and $m_{E}$
is greater than that at larger values of $\mu_q$.
At $\mu_q>0.3$~GeV the ratio $ \ds { \tilde{m}_E(\mu_q)\over m_E(\mu_q) } = \eta_E(\mu_q)$
depends only weakly on $\mu_q$: the fit of $\eta_E(\mu_q)$ to a constant 
gives $\bar \eta_E=1.6(1)$, $\ds {\chi^2 \over N_{d.o.f}} =0.51$ at $N_{d.o.f}=9$ 
and the corresponding $p-$value equals to 0.87.
One can see that $\tilde{m}_{E}$ and $m_{E}$ show a qualitatively similar
dependence on $\mu_q$. 

In the left panel of Fig.\ref{fig:both_proper_n_Linde_GeV}
we also show the function
\beq\label{eq:ff_ME_vs_muq}
\tilde{m}_E \simeq c_0 + c_2 \mu_q^2
\eeq
fitted to our values of $\tilde{m_E}$ with parameters
$c_0=0.74(3)$~GeV, $c_2=0.57(6)$~GeV$^{-1}$ and $\ds {\chi^2\over N_{d.o.f}}=1.59$.

\begin{figure}[tbh]
\centering
\vspace*{-25mm}
\includegraphics[width=10cm]{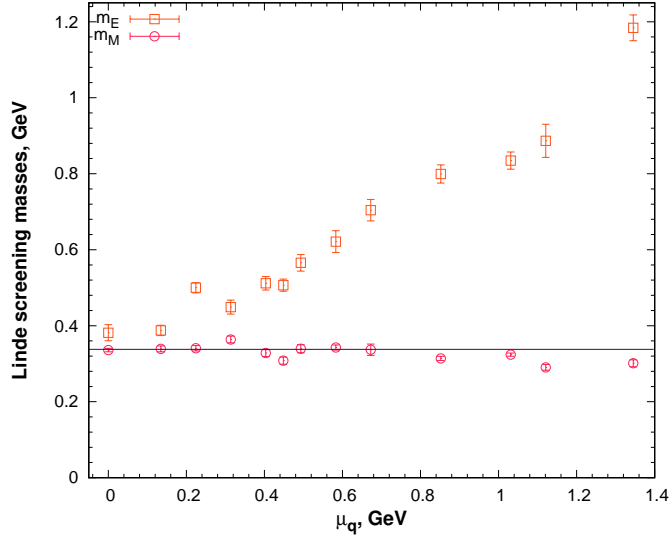}
\vspace*{-32mm}
\caption{Chromoelectric $m_E$ and chromomagnetic $m_M$ Linde screening masses 
as functions of $\mu_q$. Horizontal line results from the fit of the 
data on the Linde mass to a constant over the range $0<p<0.75$~GeV:
it is seen that at higher $\mu_q$ the Linde chromomagnetic screening mass 
tends to decrease, in contrast to the results obtained in~\cite{Boz:2018crd}.
}
\label{fig:me_mm_Linde_GeV_b}
\end{figure}

As in the chromoelectric case, the chromomagnetic ratio
$\ds \eta_M(\mu_q) =  {\tilde{m}_M(\mu_q)\over m_M(\mu_q)}$
can be fitted to a constant $\bar \eta_M = 2.1(1)$ with $\ds {\chi^2 \over N_{d.o.f}} =2.08$ 
at $N_{d.o.f}=12$ and the corresponding $p-$value equals to $0.015$.
Thus one can hardly draw a definite conclusion on 
the equivalence between chromomagnetic proper and Linde screening masses
see Fig.\ref{fig:both_proper_n_Linde_GeV}, right panel.
Moreover, as was mentioned above, 
discarding zero momentum we lose most information on the infrared behavior
of the transverse propagator. For this reason, the proper magnetic screening 
mass can hardly be reliably extracted from our data. The dependence of the chromomagnetic Linde screening mass on $\mu_q$ is shown in greater detail in Fig.\ref{fig:me_mm_Linde_GeV_b}.
together with the chromoelectric Linde screening mass.
Our results on the dependence of Linde screening masses on $\mu_q$ 
are in sharp disagreement with the results of  Ref.~\cite{Boz:2018crd}.
It was found in Ref.~\cite{Boz:2018crd} that at $a=0.138$~fm 
$m_M$ increases by some 20\%  when $\mu_q$ increases from 0 to 1.2~GeV and much faster growth was found at $a=0.186$~fm. 
In opposite, we observe a trend to decreasing of the magnetic Linde screening mass with increasing $\mu_q$. 
The chromoelectric screening mass in Ref.\cite{Boz:2018crd}
increases with $\mu_q$ at $a=0.186$~fm
and fluctuates about a constant on a finer lattice with $a=0.138$~fm.
We find that on our lattices with much smaller lattice spacing $a=0.044$~fm $m_E$ increases fast and this growth can be described by $\mu_q^2$ behavior predicted by the perturbation theory.  
From the results in Ref.\cite{Boz:2018crd} it follows
that the chromoelectric and chromomagnetic 
screening masses come close to each other at all values of $\mu_q$, 
whereas we find that they coincide only at $\mu_q=0$ and come apart
from each other as $\mu_q$ increases.
Thus lattices with spacing $a>0.13$~fm used in Ref.\cite{Boz:2018crd}
might be not sufficiently fine for the studies of screening masses. The reason may stem from the fact that the condition $\ds \mu_q <\!\!< {1 \over a}$
does not hold at large values of $\mu_q$ on such rough lattices.

\section{Perturbative behavior at high momenta and chemical potentials} 

At sufficiently high momenta it is natural to expect the RG-modified
perturbative behavior of the gluon propagator
at all values of $\mu_q$.

\begin{figure}[tbh]
\vspace*{-17mm}\hspace{-8mm}
\includegraphics[width=8.4cm]{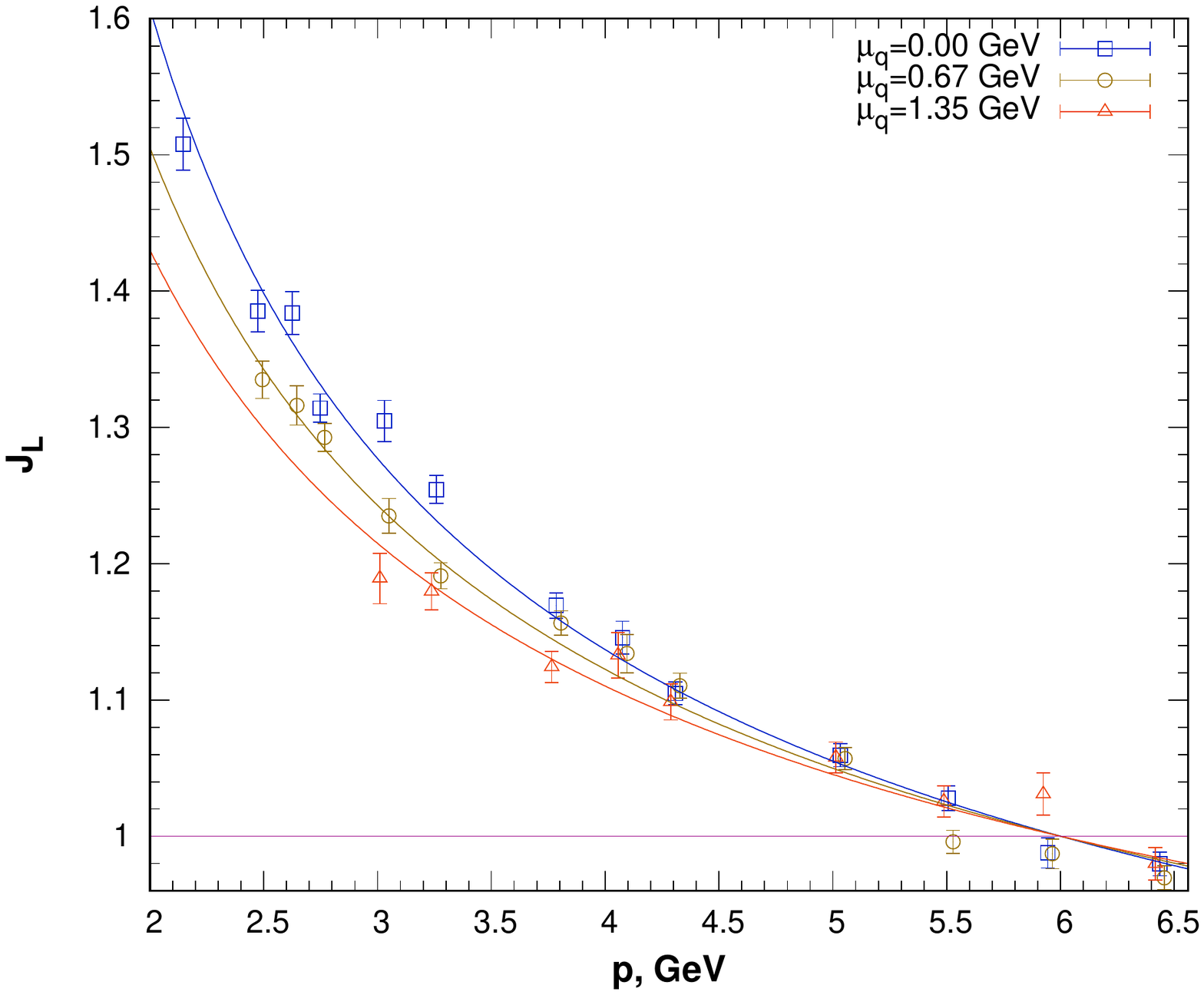}
\hspace{-6mm}
\includegraphics[width=8.4cm]{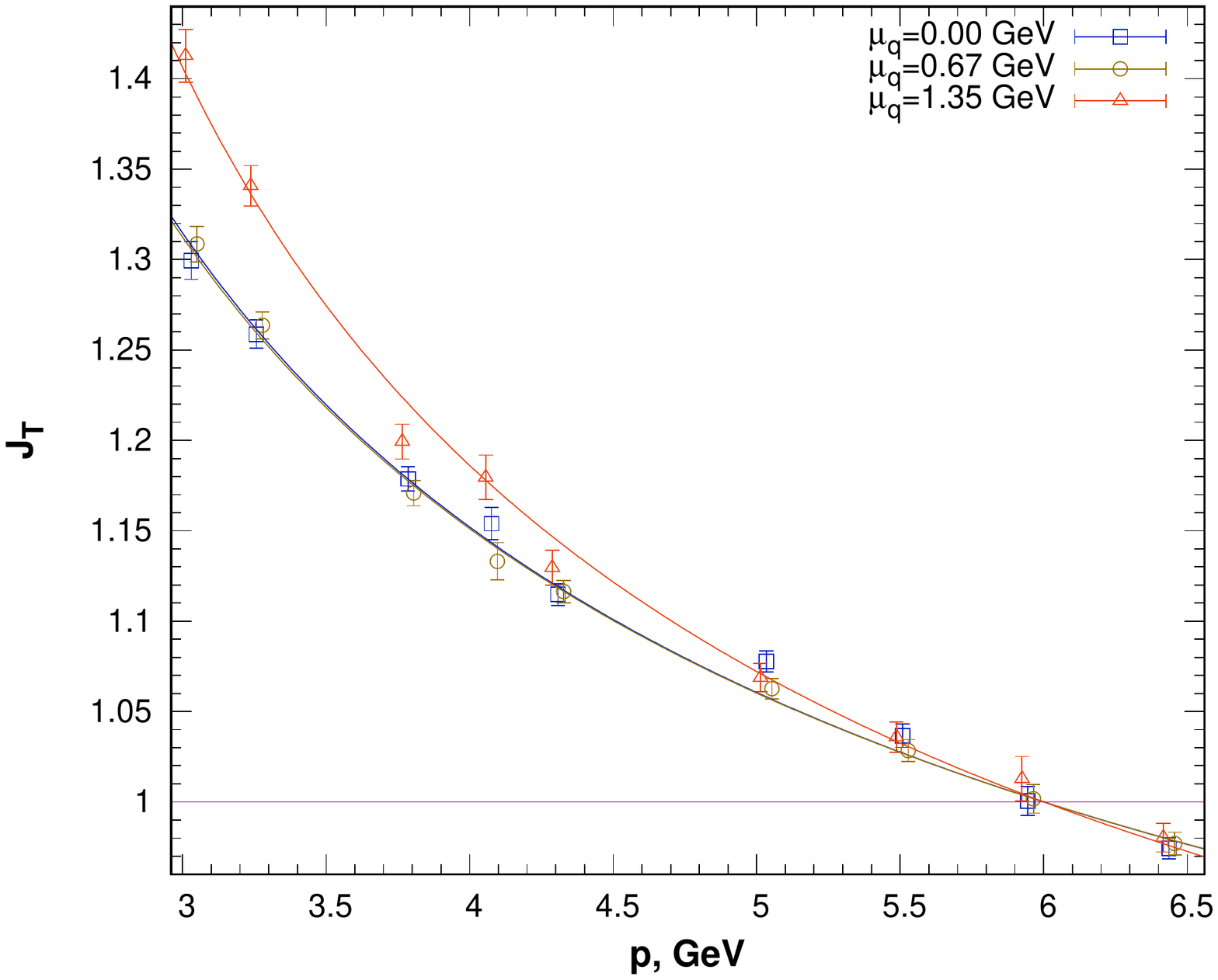}
\vspace*{-27mm}
\caption{Longitudinal (left panel) and transverse (right panel) dressing functions at various values of $\mu_q$.
Curves are obtained with the fit function (\ref{eq:PT_fit_pormula_for_J}). }
\label{fig:Dress_both_PT_b}
\end{figure}

In the one loop approximation, the 
asymptotic behavior of the gluon dressing function 
$J(p)=D(p) p^2$ has the form
\beq
\lim_{p\to \infty; g=const} J(p;g) \simeq \left[ {\ln\left({p^2\over \Lambda^2} \right) \over \ln\left({\kappa^2\over \Lambda^2}\right)} \right]^{-\; c/(2b)},
\eeq
$c$ and $b$ are the coefficients of the RG functions,
\beq
\beta(g) \simeq -bg^3, \qquad \gamma(g)\simeq -cg^2\; 
\eeq
and $\kappa$ is the normalization point.
In the Landau-gauge $SU(N_c)$ theories with $N_F$ flavors 
\cite{Politzer:1974fr} we arrive at
\beq
{c\over 2b}={13N_c-4N_F \over 2(11 N_c-2N_F)} ={1\over 2}\, .
\eeq
Thus we fit our data to the function 
\beq\label{eq:PT_fit_pormula_for_J}
J_{PT}(p) = \left[ {\ln\left({p^2\over \Lambda^2} \right) \over \ln\left({\kappa_0^2 \over \Lambda^2}\right)} \right]^{-\;0.5},
\eeq
where $\kappa_0=6$~GeV, over the domain $p>p_{cut}$. The results are shown 
in Fig.~\ref{fig:Dress_both_PT_b}.

Goodness-of-fit is decreased by the effects of $O(3)$ symmetry breaking,
however, we do not perform a systematic study of these effects
assuming that making use of the asymptotic standard error 
in the fitting  parameter $\Lambda $ takes these effects into
account. 

Each value of the cutoff momentum $p_{cut}$ is chosen 
so that (i) smaller values of $p_{cut}$ 
result in a substantial decrease of the respective $p$-value and
(ii) greater values of $p_{cut}$ 
give no significant increase of the respective $p$-value.
Thus we conclude that a domain of high momenta, where 
the longitudinal and transverse propagators can be described 
by the perturbatively motivated fit formula (\ref{eq:PT_fit_pormula_for_J}),
does exist for each value of $\mu_q$. 
In the transverse case, this domain is bounded from below by the cutoff momentum
$p_{cut}=2.9$~GeV uniformly on $\mu_q$.
In the longitudinal case, the cutoff momenta can be roughly approximated by the formula 
\beq\label{eq:cutoff_momentum}
p_{cut} = 1.8 \mbox{GeV} + 1.0 \mu_q  \;.
\eeq

\begin{figure}[tbh]
\centering
\vspace*{-17mm}
\includegraphics[width=8.0cm]{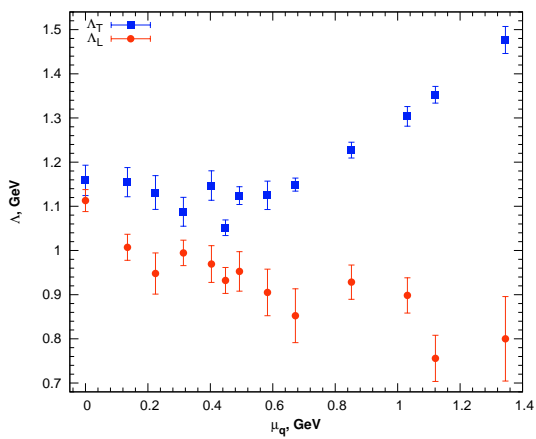}
\vspace*{-29mm}
\caption{The parameter $\Lambda$ from formula 
 (\ref{eq:PT_fit_pormula_for_J}) for the transverse and longitudinal
  dressing functions}
\label{fig:Lambda_vs_mu}
\end{figure}

The dependence of the resulting parameters on $\mu_q$ is shown in Fig.\ref{fig:Lambda_vs_mu}. 
$\Lambda_L$ and $\Lambda_T$ designate the parameter 
$\Lambda$ determined from the fit 
to $J_L$ and $J_T$, respectively.
$\Lambda_L$ gradually decreases with increasing $\mu_q$,
whereas $\Lambda_T=$const at $\mu_q<\mu_q^b \sim 700\div 800$~MeV and 
shows a linear dependence on $\mu_q$,
\beq\label{eq:LambdaT_vs_muq_ff}
\Lambda_T = \alpha_1 \mu_q + \alpha_0 \  \;  ,
\eeq
at $\mu_q>\mu_q^b$. 
Fit over the range $\mu_q>0.65$~GeV gives  $\alpha_0=0.831(17)$~GeV and $\alpha_0=0.468(18)$ with $\ds \chi^2/N_{d.o.f.}=0.19$.
Let us note that this sharp change in the behavior of $\Lambda_T(\mu_q)$
occurs at $\mu_q=\mu_q^b$, which is  only a little smaller than the value $\mu_q^{s}\sim 850$~MeV, where $\sigma_s$ starts to decrease 
(see Ref.\cite{Bornyakov:2017txe}). This value is also close 
to the chemical potential at which the string tension 
$\sigma$ vanishes.  
Therefore, the high-momentum behavior of $D_T$ changes   
in the deconfinement phase.

At $\mu_q>\mu_q^b$ the scale parameter $\Lambda_T$ 
depends on the chemical potential and, 
if formula (\ref{eq:LambdaT_vs_muq_ff}) holds true
in the limit $\mu_q\to \infty$, then
$$\ds 
{\ds \ln\left({p^2\over \Lambda_T^2} \right) \over \ds \ln\left({\kappa_0^2 \over \Lambda_T^2}\right)}\simeq {\ds \ln\left({p^2\over \alpha_1^2 \mu_q^2} \right) \over \ds \ln\left({\kappa_0^2 \over \alpha_1^2 \mu_q^2}\right)}\; .
$$ 
That is, at sufficiently high $\mu_q$ the scale parameter
in the expression for $J_T$ is proportional to the chemical potential, as it is expected, whereas the scale parameter in the expression for $J_L$ 
 depends only weakly on $\mu_q$. This controversial situation is very interesting and suggests further investigations.

\section{Conclusions}
We studied the gluon propagators in $N_f=2$  $SU(2)$ QCD at $T=0$ 
in the domain $0 < \mu_q < 1.4$~GeV, $ 0 < p < 6.5$~GeV. 
It was found that both longitudinal and transverse propagators 
depend on the chemical potential both at low and high momenta.

{\bf At low momenta,} we describe this dependence in terms of the 
chromoelectric ($m_E$) and chromomagnetic ($m_M$) screening masses
using two definitions: Linde screening masses $m_{E,M}$ and 
proper screening masses $\tilde m_{E,M}$. 
We found a good agreement between the two definitions
of the chromoelectric screening mass at least at $\mu_q > 0.3$~GeV.
$m_E$ increases substantially with $\mu_q$ and can be fitted  by the 
function  (\ref{eq:ff_ME_vs_muq}).

The case of the chromomagnetic screening mass is more complicated: 
we find only a rough agreement between the two definitions. 
The Linde mass $m_M$ can be evaluated more 
precisely; it depends only weakly on $\mu_q$
and can be fitted well by a constant at $\mu_q < 0.8 $~GeV. 
At higher $\mu_q$ one can see decreasing of $m_M$ 
which agrees with decreasing of $\sigma_s$. 
Results for higher values of $\mu_q$ are needed 
to decide whether $m_M$ goes to zero at large $\mu_q$
as was argued in Ref.\cite{Son:1998uk}.
In any case, the difference between $m_E$ and $m_M$
shows a substantial growth with $\mu_q$ starting at $\mu_q\approx 0.3$~GeV
(see Fig.\ref{fig:me_mm_Linde_GeV_b}).

It should be emphasized that our findings contradict to the
results of Ref.~\cite{Boz:2018crd}, where it was concluded
that {\it (i)} $m_M$ comes close to $m_E$ for all $\mu_q$ and
{\it (ii)} both screening masses depend only weakly on $\mu_q$. 

{\bf At high momenta,} we used the perturbatively motivated fit function
(\ref{eq:PT_fit_pormula_for_J}) and described $\mu_q$-dependence 
of the propagators $D_{T,L}$ in terms of the scaling parameters $\Lambda_{T,L}$
that appear in formulas like (\ref{eq:PT_fit_pormula_for_J}) for $D_T$ and $D_L$.

$\Lambda_L$ shows a slow decrease with increasing $\mu_q$,
whereas $\Lambda_T=$const at $\mu_q < 750$~MeV and 
shows a linear growth at higher values of $\mu_q$.
A sharp change in the behavior of $\Lambda_T(\mu_q)$
occurs at $\mu_q$ where the spatial string tension $\sigma_s$ peaks 
(see Ref.\cite{Bornyakov:2017txe}). \\[1mm]

{\bf Acknowledgments.} The authors are grateful to V. Braguta for useful discussions. The work was completed due to support of the Russian Foundation for Basic Research via grant 18-02-40130 mega (analysis of gluon propagators) 
and via grant 18-32-20172 (gauge fixing and Gribov copy effects analysis). 
A.A.N. acknowledges support from STFC via grant ST/P00055X/1.
The research is carried out using the equipment of the shared research facilities of HPC computing resources at Lomonosov Moscow State University,
Central Linux Cluster of the NRC "Kurchatov Institute" - IHEP (Protvino), and Linux Cluster of the NRC "Kurchatov Institute" - ITEP (Moscow).



\bibliographystyle{apsrev}

\begin{thebibliography}{35}
\expandafter\ifx\csname natexlab\endcsname\relax\def\natexlab#1{#1}\fi
\expandafter\ifx\csname bibnamefont\endcsname\relax
  \def\bibnamefont#1{#1}\fi
\expandafter\ifx\csname bibfnamefont\endcsname\relax
  \def\bibfnamefont#1{#1}\fi
\expandafter\ifx\csname citenamefont\endcsname\relax
  \def\citenamefont#1{#1}\fi
\expandafter\ifx\csname url\endcsname\relax
  \def\url#1{\texttt{#1}}\fi
\expandafter\ifx\csname urlprefix\endcsname\relax\def\urlprefix{URL }\fi
\providecommand{\bibinfo}[2]{#2}
\providecommand{\eprint}[2][]{\url{#2}}

\bibitem{Kogut:2000ek} 
  J.~B.~Kogut, M.~A.~Stephanov, D.~Toublan, J.~J.~M.~Verbaarschot and A.~Zhitnitsky,
  Nucl.\ Phys.\ B {\bf 582}, 477 (2000)
  [hep-ph/0001171].
  
\bibitem{Splittorff:2001fy} 
  K.~Splittorff, D.~Toublan and J.~J.~M.~Verbaarschot,
  Nucl.\ Phys.\ B {\bf 620}, 290 (2002)
  [hep-ph/0108040].

\bibitem{Kanazawa:2009ks} 
  T.~Kanazawa, T.~Wettig and N.~Yamamoto,
  JHEP {\bf 0908}, 003 (2009)
  [arXiv:0906.3579 [hep-ph]].
  
\bibitem{Brauner:2009gu} 
  T.~Brauner, K.~Fukushima and Y.~Hidaka,
  Phys.\ Rev.\ D {\bf 80}, 074035 (2009)
  Erratum: [Phys.\ Rev.\ D {\bf 81}, 119904 (2010)]
  [arXiv:0907.4905 [hep-ph]].
  
\bibitem{Sun:2007fc} 
  G.~f.~Sun, L.~He and P.~Zhuang,
  Phys.\ Rev.\ D {\bf 75}, 096004 (2007)
  [hep-ph/0703159].
  
\bibitem{He:2010nb} 
  L.~He,
  Phys.\ Rev.\ D {\bf 82}, 096003 (2010)
  [arXiv:1007.1920 [hep-ph]].
 
\bibitem{Strodthoff:2011tz} 
  N.~Strodthoff, B.~J.~Schaefer and L.~von Smekal,
  Phys.\ Rev.\ D {\bf 85}, 074007 (2012)
  [arXiv:1112.5401 [hep-ph]].

\bibitem{Strodthoff:2013cua} 
  N.~Strodthoff and L.~von Smekal,
  Phys.\ Lett.\ B {\bf 731}, 350 (2014)
  [arXiv:1306.2897 [hep-ph]].

\bibitem{Vanderheyden:2001gx} 
  B.~Vanderheyden and A.~D.~Jackson,
  Phys.\ Rev.\ D {\bf 64}, 074016 (2001)
  [hep-ph/0102064].
  
  
\bibitem{Kanazawa:2011tt} 
  T.~Kanazawa, T.~Wettig and N.~Yamamoto,
  JHEP {\bf 1112}, 007 (2011)
  [arXiv:1110.5858 [hep-ph]].


\bibitem{Hands:1999md} 
  S.~Hands, J.~B.~Kogut, M.~P.~Lombardo and S.~E.~Morrison,
  Nucl.\ Phys.\ B {\bf 558}, 327 (1999)
  [hep-lat/9902034].

\bibitem{Kogut:2001na} 
  J.~B.~Kogut, D.~K.~Sinclair, S.~J.~Hands and S.~E.~Morrison,
  Phys.\ Rev.\ D {\bf 64}, 094505 (2001)
  [hep-lat/0105026].

\bibitem{Kogut:2001if} 
  J.~B.~Kogut, D.~Toublan and D.~K.~Sinclair,
  Phys.\ Lett.\ B {\bf 514}, 77 (2001)
  [hep-lat/0104010].

\bibitem{Kogut:2002cm} 
  J.~B.~Kogut, D.~Toublan and D.~K.~Sinclair,
  Nucl.\ Phys.\ B {\bf 642}, 181 (2002)
  [hep-lat/0205019].

\bibitem{Braguta:2016cpw} 
  V.~V.~Braguta, E.-M.~Ilgenfritz, A.~Y.~Kotov, A.~V.~Molochkov and A.~A.~Nikolaev,
  Phys.\ Rev.\ D {\bf 94}, no. 11, 114510 (2016)
  [arXiv:1605.04090 [hep-lat]].

\bibitem{Bornyakov:2017txe} 
  V.~G.~Bornyakov, V.~V.~Braguta, E.-M.~Ilgenfritz, A.~Y.~Kotov, A.~V.~Molochkov and A.~A.~Nikolaev,
  JHEP {\bf 1803}, 161 (2018)
  [arXiv:1711.01869 [hep-lat]].

\bibitem{Astrakhantsev:2018uzd} 
  N.~Y.~Astrakhantsev, V.~G.~Bornyakov, V.~V.~Braguta, E.-M.~Ilgenfritz, A.~Y.~Kotov, A.~A.~Nikolaev and A.~Rothkopf,
  JHEP {\bf 1905}, 171 (2019)
  doi:10.1007/JHEP05(2019)171
  [arXiv:1808.06466 [hep-lat]].
  
\bibitem{Wilhelm:2019fvp} 
  J.~Wilhelm, L.~Holicki, D.~Smith, B.~Wellegehausen and L.~von Smekal,
  Phys.\ Rev.\ D {\bf 100}, no. 11, 114507 (2019)
  doi:10.1103/PhysRevD.100.114507
  [arXiv:1910.04495 [hep-lat]].
  
\bibitem{Nakamura:1984uz} 
  A.~Nakamura,
  Phys.\ Lett.\  {\bf 149B}, 391 (1984).
  doi:10.1016/0370-2693(84)90430-1
  
\bibitem{Hands:2006ve} 
  S.~Hands, S.~Kim and J.~I.~Skullerud,
  Eur.\ Phys.\ J.\ C {\bf 48}, 193 (2006)
  doi:10.1140/epjc/s2006-02621-8
  [hep-lat/0604004].

\bibitem{Hands:2010gd} 
  S.~Hands, S.~Kim and J.~I.~Skullerud,
  Phys.\ Rev.\ D {\bf 81}, 091502 (2010)
  doi:10.1103/PhysRevD.81.091502
  [arXiv:1001.1682 [hep-lat]].

\bibitem{Hands:2011ye} 
  S.~Hands, P.~Kenny, S.~Kim and J.~I.~Skullerud,
  Eur.\ Phys.\ J.\ A {\bf 47}, 60 (2011)
  doi:10.1140/epja/i2011-11060-1
  [arXiv:1101.4961 [hep-lat]].

\bibitem{Cotter:2012mb} 
  S.~Cotter, P.~Giudice, S.~Hands and J.~I.~Skullerud,
  Phys.\ Rev.\ D {\bf 87}, no. 3, 034507 (2013)
  doi:10.1103/PhysRevD.87.034507
  [arXiv:1210.4496 [hep-lat]].

 
\bibitem{Boz:2018crd} 
  T.~Boz, O.~Hajizadeh, A.~Maas and J.~I.~Skullerud,
  Phys.\ Rev.\ D {\bf 99}, no. 7, 074514 (2019)
  [arXiv:1812.08517 [hep-lat]].

\bibitem{Weisz:1982zw} 
  P.~Weisz,
  Nucl.\ Phys.\ B {\bf 212}, 1 (1983).

\bibitem{Bazavov:2011nk} 
  A.~Bazavov {\it et al.},
  Phys.\ Rev.\ D {\bf 85}, 054503 (2012)
  [arXiv:1111.1710 [hep-lat]].


\bibitem{Mandula:1987rh} 
  J.~E.~Mandula and M.~Ogilvie,
  Phys.\ Lett.\ B {\bf 185}, 127 (1987).
  
  
\bibitem{Bornyakov:2009ug} 
  V.~G.~Bornyakov, V.~K.~Mitrjushkin and M.~Muller-Preussker,
  Phys.\ Rev.\ D {\bf 81}, 054503 (2010)
  [arXiv:0912.4475 [hep-lat]].
  
\bibitem{Kapusta:2006pm} 
  J.~I.~Kapusta and C.~Gale,
  Finite-temperature field theory: Principles and applications,
  Cambridge University Press, 2011, doi:10.1017/CBO9780511535130

\bibitem{Bornyakov:2010nc} 
  V.~G.~Bornyakov and V.~K.~Mitrjushkin,
  Phys.\ Rev.\ D {\bf 84}, 094503 (2011)
  [arXiv:1011.4790 [hep-lat]].
  
\bibitem{Dudal:2018cli} 
  D.~Dudal, O.~Oliveira and P.~J.~Silva,
  Annals Phys.\  {\bf 397}, 351 (2018)
  [arXiv:1803.02281 [hep-lat]].

\bibitem{Linde:1980ts} 
  A.~D.~Linde,
  Phys.\ Lett.\  {\bf 96B}, 289 (1980).

\bibitem{Maas:2011se} 
  A.~Maas,
  Phys.\ Rept.\  {\bf 524}, 203 (2013)
  [arXiv:1106.3942 [hep-ph]].

\bibitem{Politzer:1974fr} 
  H.~D.~Politzer,
  Phys.\ Rept.\  {\bf 14}, 129 (1974).

\bibitem{Son:1998uk}
      D.~T.~Son, 
      Phys.\ Rev.\ {\bf D59}, 094019 (1999), 
     [arXiv:[hep-ph/9812287]].
 
\end{thebibliography}

\end{document}